\documentclass[10pt,letterpaper]{article}
\usepackage[english]{babel}
\usepackage{graphicx}

\newcommand{\alt}{\mathbin{\lower 3pt\hbox
   {$\rlap{\raise 5pt\hbox{$\char'074$}}\mathchar"7218$}}}
\newcommand{\agt}{\mathbin{\lower 3pt\hbox
   {$\rlap{\raise 5pt\hbox{$\char'076$}}\mathchar"7218$}}}

\begin{document}
\setcounter{footnote}{0}
\setcounter{equation}{0}
\setcounter{figure}{0}
\setcounter{table}{0}
\vspace*{5mm}

\begin{center}
{\large\bf Interpretation of high-dimensional \\
numerical results for Anderson transition }

\vspace{4mm}
I. M. Suslov \\
P.L.Kapitza Institute for Physical Problems,
\\ 119334 Moscow, Russia \\
E-mail: suslov@kapitza.ras.ru
\\
\vspace{5mm}
\end{center}

\begin{center}
\begin{minipage}{135mm}
{\large\bf Abstract } \\
Existence of the upper critical dimension  $d_{c2}=4$
for the Anderson transition is a rigorous consequence of
the Bogoliubov theorem on renormalizability of  $\phi^4$
theory. For dimensions $d\ge 4$, one-parameter scaling does
not hold, and all existent numerical data should be
reinterpreted. These data are exhausted by results for
 $d=4,\,5$ from scaling in quasi-one-dimensional systems,
 and results for  $d=4,\,5,\,6$ from level statistics.
All these data are compatible with the theoretical scaling
dependencies obtained from self-consistent theory of localization
by Vollhardt and W${\rm {\ddot o}}$lfle. The critical discussion
is given for a widespread point of view that  $d_{c2}=\infty$.
\end{minipage}
\end{center}
\vspace{5mm}

\twocolumn


\vspace{6mm}
\begin{center}
{\bf 1. Introduction}
\end{center}

The main defect of the current literature on the Anderson
transition is the ignorance of the upper critical dimension
$d_{c2}=4$, which is a rigorous consequence of the Bogoliubov
theorem on renormalizability of $\phi^4$ theory \cite{1,2}. The
problem of the Anderson transition can be reduced (in the
mathematically exact manner) to one of the variants of  $\phi^4$
theory \cite{3,4,5,6}\,\footnote{\,Specifically, to a problem of
two zero-component interacting fields \cite{4,5}. The arguments on
the deficiency  of the replica method  \cite{7,8} are inessential
in the given context, since the analysis of renormalizability can
be carried out on the diagrammatic level. Each diagram of the
disordered systems theory can be obtained from a certain diagram
for  $\phi^4$ theory by a simple replacement of symbols
\cite{3}.}, which is nonrenormalizable for  space dimensions
$d>4$. Therefore, the cut-off momentum  $\Lambda$, corresponding
to the atomic length scale $a$, cannot be excluded from results.
It rules out the existence of one-parameter scaling \cite{9},
according to which  the correlation radius $\xi$ is the only
essential length scale. In the latter case, any dimensionless
quantity  $Q$, related to a finite system of size $L$, can be
written as a function of ratio  $L/\xi$,
$$
Q=F(L/\xi)\,,
\eqno(1)
$$
which is a base for all numerical algorithms. The study of  $Q$ as
a function of $L$ and a distance to the transition  $\tau$ allows
to determine the critical exponent $\nu$  of the correlation
length  ($\xi\sim |\tau|^{-\nu}$). Indeed, if two $L$-dependencies
for $\tau=\tau_1$ and $\tau=\tau_2$ are calculated, then the scale
transformation allows to determine the ratio of two correlation
lengths. Producing this procedure for a  succession $\tau_1$,
$\tau_2$ $\tau_3,\,\ldots$, one can determine $\xi(\tau_i)$ apart
from numerical factor.

Relation (1) is invalid for $d>4$, and the more general form
should be used,
$$
Q=F(L/\xi, L/a) \,.
\eqno(2)
$$
For  $d=4$, a situation is more complicated and needs the
additional study; in fact, the existence of logarithmic factors
like $\ln(L/a)$ leads to the relation of type (2). The latter can
be reduced to a function of one argument, if an appropriate choice
of the scaling variables is made \cite{14,15}.

Currently, there exist the following results for the Anderson
transition in high dimensions: the data by Markos for
$d=4$, $d=5$ \cite{10,11}, obtained from scaling in
quasi-one-dimensional systems; the data by Zharekeshev and
Kramer for  $d=4$ \cite{12}, and those by Garcia-Garcia and
Cuevas for  $d=5$, $d=6$ \cite{13}, obtained from the
level statistics. All these results are based on  relation
(1) and need reinterpretation due to above arguments.

Below  (Secs.\,3, 4) these results are compared with the modified
scaling for high dimensions obtained in  \cite{14,15} from
self-consistent theory of localization by Vollhardt and W${\rm
{\ddot o}}$lfle \cite{16}. The latter gives correct values of the
upper critical dimension  $d_{c2}=4$ and the exponent  $\nu=1/2$
for  $d>d_{c2}$, and at least is of interest as a possible
scenario.  According to certain arguments  \cite{17,18}, the
Vollhardt and W${\rm {\ddot o}}$lfle theory predicts the exact
critical behavior, and a lot of numerical results can be matched
with it \cite{14,15,19,20}. The present paper supports the same
tendency: all indicated numerical data  \cite{11,12,13} can be
matched with  theoretical scaling dependencies. As a rule, the
"experimental" points lie on the quasi-linear portions of the
scaling curves, and was interpreted as dependence $L^{1/\nu}$ with
$\nu\approx 1$ in the original papers. The actual critical
behavior suggests   $\nu=1/2$, but the corresponding parts of the
scaling dependencies are difficult for study in numerical
experiments due to their restricted accuracy.

The widespread viewpoint that  $d_{c2}=\infty$ \cite{21a}--\cite{25},
is discussed in the next section.
\begin{figure*}
\centerline{\includegraphics[width=5.1 in]{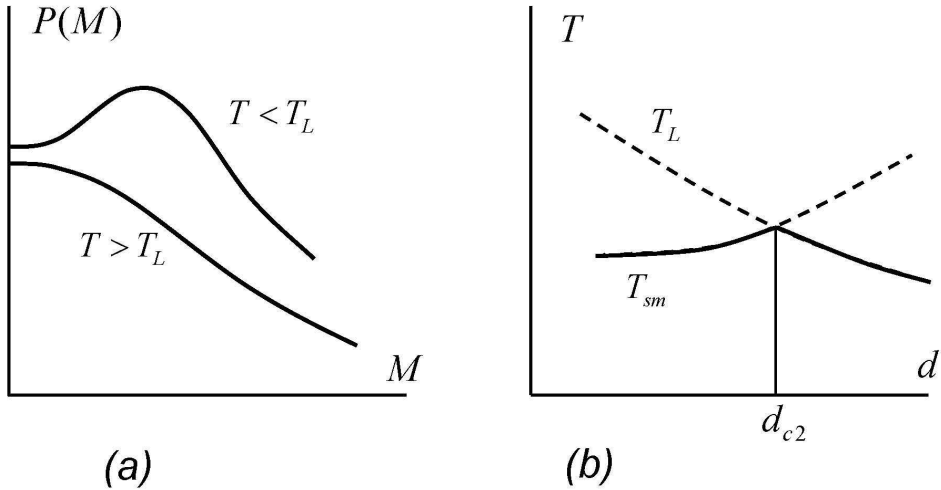}} \caption{ (a)
One can use the sigma-model concept only in the case when the
distribution function  $P(M)$ of the modulus of ${\bf M}$ has a
maximum at finite  $M$. Such a maximum arises as a result of the
mean-field type transition  in the spirit of the Landau theory.
(b) The temperature $T_L$ at which a sigma-model arises and the
critical point $T_{sm}$ according to the "sigma-model scenario" as
functions of space dimensionality $d$. Solid and dashed lines show
the real and fictitious phase transitions.    } \label{fig1}
\end{figure*}

\begin{center}
{\bf 2. Sigma-models and  $d_{c2}$.}
\end{center}

A hypothesis that  $d_{c2}=\infty$ is based on the following
arguments:
\vspace{2mm}

(a) In the approach based on the use of the sigma-models
\cite{100}, there are no indications on existence of the
special dimensionality in the interval
$2<d<\infty$ \cite{21a}.

(b) Results  $s=\infty$, $\nu=1/2$ for $d=\infty$  \cite{21,22}
($s$ is the critical exponent of conductivity) show
validity of the Wegner relation  $s=\nu(d-2)$ for  $d=\infty$,
and one can expect that this relation (and, consequently,
the one-parameter scaling picture  \cite{9}) is valid for
all  $d>2$.  \vspace{2mm}

We do not question results $s=\infty$, $\nu=1/2$ for the
infinite-dimensional sigma-model, but there is a problem of their
correspondence with the initial disordered system. For electrons
in the random potential, derivation of sigma-models is
substantiated only for dimensions  $d=2+\epsilon$ with
$\epsilon\ll 1$; qualitatively, it can be extended to  $\epsilon
\sim 1$ but not to $d\gg 1$. Extension of sigma-models to higher
dimensions is based on the artificial construction corresponding
to a system of weakly connected metallic granules \cite{21a}. In
each granule, only the zero Fourier component of the matrix field
$Q$ is taken into account, while connections between granules are
supposed to produce only slow variation of $Q$. The possibility
that a coupling between granules leads to induction of higher
Fourier components and a practical destruction of the sigma-model
is not considered, while such situation looks rather probable from
a standpoint of spatially homogeneous systems\,\footnote{\,In our
opinion, it is practically evident. If field $Q$ is indeed slowly
varying, then it remains almost constant inside the block composed
of several granules; it means validity of the Wigner--Dyson
statistics for this block \cite{26}. In fact, for the strength of
coupling corresponding to the Anderson transition, hybridization
of the block eigenfunctions is not complete (i.e. with equal
weights, as in a metallic phase) but partial (which is typical for
the critical region). Hence, the level statistics for the composed
system of several granules will be essentially different from the
Wigner--Dyson one. }.

Let  explain a situation on the example of the vector sigma-model.
If a ferromagnet is described in terms of $\phi^4$ theory, then
the magnetic moment ${\bf M}$ of a finite block can be considered
as the Heisenberg spin with a certain fluctuation of its modulus
$M$. Neglecting of such longitudinal fluctuations  (which can be
rigorously justified in dimensions  $d=2+\epsilon$) by definition
corresponds to a sigma-model. For  $d=3$, the longitudinal
fluctuations  have no qualitative effect and the sigma-model looks
applicable\,\footnote{\,In fact, it is possible to show  (see
\cite{27}, Sec.3.1) that equivalence of the sigma-model and
$\phi^4$ theory in the sense of the critical behavior takes place
for $d<4$.}. However, when $d$ approaches to  4, the longitudinal
fluctuations become anomalously soft and this is the origin of
the upper critical dimension. If longitudinal fluctuations are
artificially suppressed (which is a case in the sigma-models),
then it can lead to elimination of $d_{c2}$.

In fact, one can use the sigma-model concept only in the case,
when the distribution function  $P(M)$ of the modulus of  ${\bf
M}$ has a maximum at finite $M$ (Fig.1,a). With a decrease of
temperature,  such a maximum arises in the result of the mean
field type  transition in the spirit of the Landau theory.
However, the corresponding temperature $T_L$ does not signify  the
actual phase transition, since the transverse fluctuations of
${\bf M}$ destroy the long range order. Long range correlations
for the transversal fluctuations arise at the lower temperature
$T_{sm}$.  Therefore, with a decrease of temperature, firstly (at
point $T_L$) a sigma-model arises, and secondly (at  point
$T_{sm}$) the phase transition takes place according to the
"sigma-model scenario". One can imagine that with change of $d$,
dependencies
 $T_L(d)$ and $T_{sm}(d)$ intersect at  point  $d_{c2}$
 (Fig.1,b). Hence, for  $d>d_{c2}$ the phase transition
occurs at the same point $T_L$ where a sigma-model arises,
i.e. according to the "Landau scenario".
\begin{figure}
\centerline{\includegraphics[width=3.1 in]{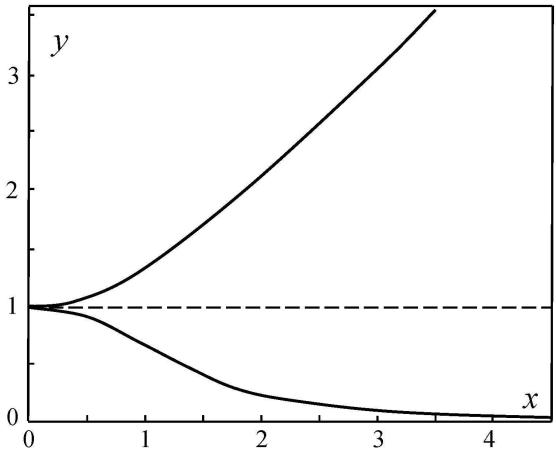}}
\caption{Scaling function  $y(x)$, corresponding to
high dimensions,  for  an algorithm based on the
use of the auxiliary quasi-one-dimensional systems. } \label{fig2}
\end{figure}

One can see that the "sigma-model transition" (at  point
$T_{sm}$) is not realized for $d>d_{c2}$, if the properties of
the sigma-model correspond to the properties of the real system.
However, such transition can exist (as a theoretical
construction) and correspond to  the analytical continuation from
lower dimensions, if a sigma-model is introduced artificially.
In our opinion, exactly such situation takes place in the
Anderson transition theory: the results $s=\infty$, $\nu=1/2$
for $d=\infty$ correspond to the formal sigma-model
and not to the initial disordered
system\,\footnote{\,According to  \cite{18}, the high-dimensional
sigma-model is unstable to small perturbations of the general
form, as a consequence of the unusual result
$\epsilon(0,0)\sim \xi$ for the dielectric constant
$\epsilon(\omega,q)$; its difference from the natural
result $\epsilon(0,0)\sim \xi^2$ indicates the existence
of the special dimension.
It should be noted, that a physical sense of the upper
critical dimension $d_{c2}=4$ can be completely clarified
for the problem of the density of states \cite{6}.}.
Direct analysis of the Bete lattice  (without use of
sigma-models) gives results $s=1$, $\nu=1/2$ \cite{28,29}, in
correspondence with the Vollhardt and W${\rm {\ddot o}}$lfle
theory.

In addition, let  discuss the paper \cite{25} where a conclusion
that $d_{c2}=\infty$ is drawn from the modification of
self-consistent theory. This paper is based on the wrong idea that
the  $L$ dependence for the diffusion constant in the critical
point, $D_L\sim L^{2-d}$, signifies the existence of the momentum
dependence  $D(q)\sim q^{d-2}$. In fact, the dependence on $L$
arises not only from spatial, but also from the temporal
dispersion, and for a given function  $D(\omega, q)$ is determined
by relation
$$
D_L\sim D\left( D_L/L^2, L^{-1} \right) \,.
\eqno(3)
$$
If a power law dependence on $\omega$ and $q$ is accepted,
then  a combination
$$
D(\omega, q)  \sim \omega^{\eta/d} q^{d-2-\eta}
\eqno(4)
$$
provides the correct behavior  $D_L\sim L^{2-d}$ for any
value of the exponent  $\eta$ \cite{30}.  If one repeat the
construction of \cite{25} with an arbitrary value of  $\eta$,
then it is easy to test that the Wegner relation
$s=\nu(d-2)$ is valid (for $d<d_{c2}$) only for $\eta=d-2$, i.e.
in the absence of the spatial dispersion\,\footnote{\,Absence of
the spatial dispersion of  $D(\omega,q)$ is obtained in \cite{18}
by a detailed analysis. Arguments relating the exponent  $\eta$
with multifractality of wave functions \cite{30}, are logically
 defective
\cite{31}.  }. For the choice  $\eta=0$ made in \cite{25}, the
Wegner relation is violated and the main argument of this paper
(agreement with numerics) becomes fictitious, since all numerical
results are based on the one-parameter scaling \cite{9}.

\begin{center}
{\bf 3. Quasi-one-dimensional systems.}
\end{center}

One of the popular numerical algorithms is based on
consideration of the auxiliary quasi-one-dimensional
systems \cite{32}, whose  correlation length  $\xi_{1D}$
is always finite. As a scaling parameter one use the
quantity \cite{10,14}
$$
z_1=L/\xi_{1D} \,.
\eqno(5)
$$
 In numerical experiments,
$\xi_{1D}$ is estimated through the minimal Lyapunov
exponent \cite{32,33}, while the scaling relation of type (1) is
postulated for  $z_1$. In high dimensions such relation is
invalid and one should use the modified scaling suggested in
 \cite{14}. The theoretical scaling function  $y(x)$ is
determined by equation
$$
\pm \, x^2= \frac{1}{y} -y^2   \,,
\eqno(6)
$$
where variables $y$ are $x$ are defined as
$$
y=\frac{L}{\xi_{1D}} \left(\frac{L}{a} \right)^{(d-4)/3},\qquad
x=\frac{L}{\xi} \left(\frac{L}{a} \right)^{(d-4)/3},
\eqno(7)
$$
for $d>4$, and
$$
y=\frac{ L}{\xi_{1D}} \left[\ln(L/a)\right]^{1/3}
\,,\qquad
x=\frac{ L }{\xi} \frac{\left[\ln(\xi/a)\right]^{1/2} }
  {\left[\ln(L/a)\right]^{1/6}}     \,,
\eqno(8)
$$
for $d=4$. Dependence  $y(x)$ consists of two branches and
is shown in Fig.2. It is clear from (6--8) that the usual
scaling constructions are possible, if the quantity $y$
is considered as a function of the "modified length"
 $\mu(L)=L^{(d-1)/3}$ ($d>4$) or  $\mu(L)=L[\ln(L/a)]^{-1/6}$
 ($d=4$).

Fig.3,a illustrates the numerical data by Markos for $d=4$
extracted from Fig.61 of  paper \cite{10} and presented as
 $z_1 (\ln L)^{1/3}$  versus $\mu(L)$. The constant limit is
 realized for  $W=33$, which gives the estimate of the
 critical point somewhat different in comparison with
 $W_c=34.3$ in \cite{10}. Accepting $y(W,L)-y(33,L)$ as
  $y-y_c$, one can put all numerical data on the theoretical
  scaling curve by the change of the scale along the horizontal
 axis  (Fig.3,b), if the common scale along the $y$ axis
 is chosen in appropriate manner.
\begin{figure*}
\centerline{\includegraphics[width=6.6 in]{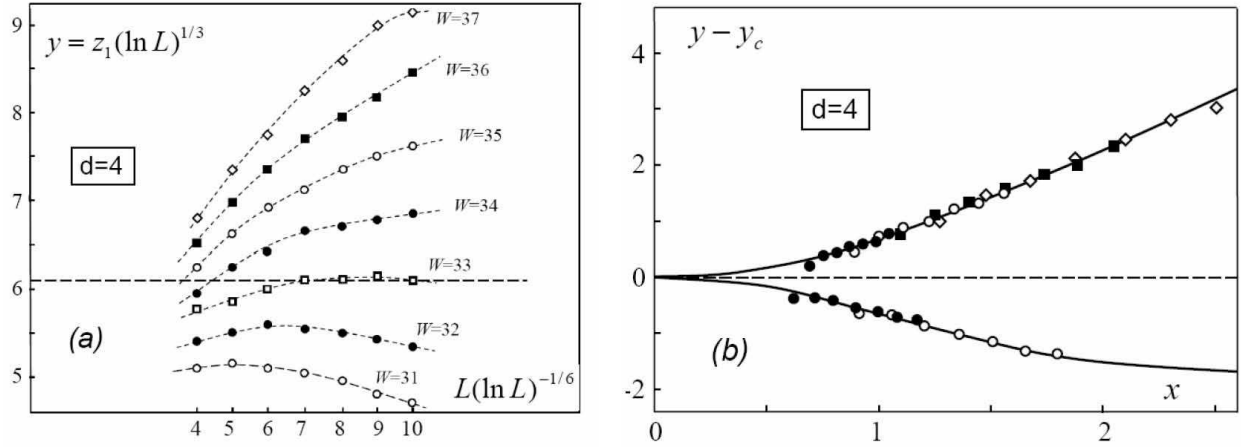}}
\caption{(a) Numerical data by Markos for  $d=4$
(quasi-one-dimensional systems) extracted from Fig.61 of
paper \cite{10}; figures near the horizontal axis show the
corresponding value of $L$. (b) Comparison with the
theoretical scaling dependence. }
\label{fig3}
\end{figure*}
\begin{figure*}
\centerline{\includegraphics[width=6.6 in]{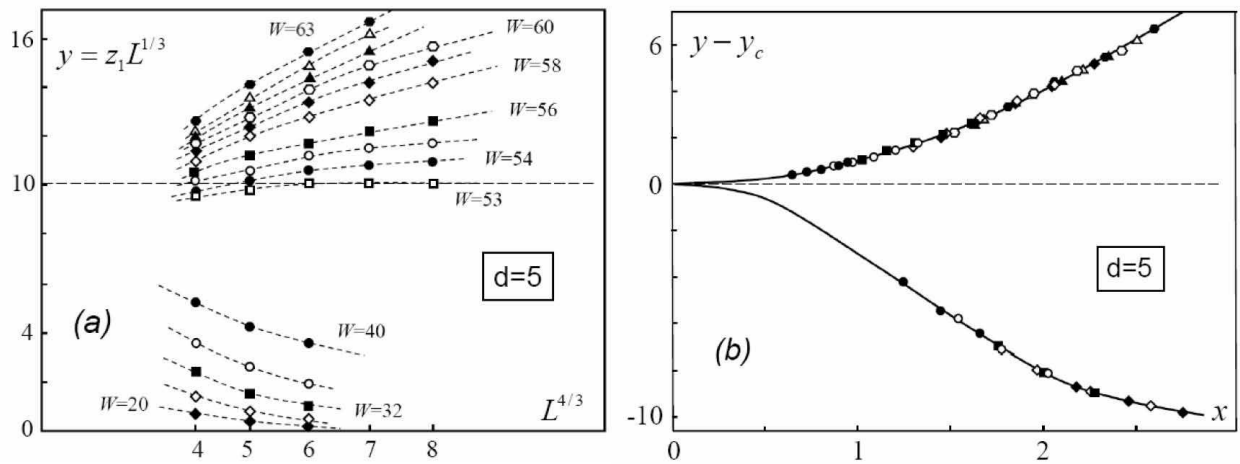}}
\caption{Numerical data by Markos for  $d=5$
(quasi-one-dimensional systems) extracted from Fig.61 of paper
\cite{10} and Figs.4,\,5 of paper \cite{11}  (a), and their
comparison with the theoretical scaling dependence (b). }
\label{fig4}
\end{figure*}

Fig.4,a  illustrates the numerical data by Markos for  $d=5$
extracted from Fig.61 of \cite{10} and Figs.4,5 of \cite{11},
presented as  $z_1 L^{1/3}$  versus $L^{4/3}$. The change of the
treatment procedure led to the essential shift of the critical
point, from  $W_c=57.3$ \cite{10} to  $W_c=53$, and made it close
to the estimate  $W_c=51.4$ of paper  \cite{13}, obtained from the
level statistics (see Sec.\,4). Accepting $y(W,L)-y(53,L)$ as
$y-y_c$, one can put all experimental points on the theoretical
scaling curve (Fig.4,b).

In both cases, the main body of data lies on quasi-linear portions
$y\sim x$  of the scaling curves and corresponds to dependencies
$z_1\sim L (\ln L)^{1/3}$ for $d=4$ and $z_1\sim L$ for $d=5$,
which were interpreted in  \cite{10} as $z_1\sim L^{1/\nu}$ with
$\nu\approx 1$. In fact, these data are consistent with predictions
of the  Vollhardt and W${\rm {\ddot o}}$lfle  theory, which
gives  $\nu=1/2$; the corresponding dependence $y-y_c\sim x^2$ is
valid only for small deviations from the critical point, which are
comparable with the scattering of  experimental points.

To conclude the section, let discuss the technical moment related
with a choice of the scaling procedure. The scaling constructions
can be carried out in the usual or logarithmic coordinates, which
is absolutely identical in the case of the rigorous scaling. In
the actual situation, the logarithmic scaling may not provide a
sufficiently smooth matching of two "pieces" of the measured
dependence, since it rigidly fix the origin of the $L$ axis.
Scaling in the usual coordinates allows a more smooth matching of
pieces due to small shifts along the horizontal axis. Such shifts
should be absent in the case of exact scaling, but practically
they arise due to scaling corrections. For example, the  structure
of scaling corrections to relation  (1) has a following form for
small  $\tau$ \cite{14}
$$
y-y_c= \tau \left\{ A_0 L^{1/\nu} + A_1 L^{\omega_1} + A_2
L^{\omega_2} +\ldots\right\}+
$$
$$
+\left\{B_1 L^{-\alpha_1} + B_2 L^{-\alpha_2} +\ldots
\right\}\,,
\eqno(9)
$$
where  $1/\nu>\omega_1>\omega_2>\ldots$,
$\alpha_1<\alpha_2<\ldots$. In the accepted interpretation of
$y-y_c$  as $y(W,L)-y(W_c,L)$, the term in the second brackets is
excluded from consideration. The expression in the first brackets
is dominated by  $L^{1/\nu}$ for large $L$, while other terms are
dominant for small  $L$: effectively, it shifts the origin of the
$L$ axis. With variable  $L/\xi$ used instead of $L$, such shift
becomes  $\tau$-dependent. If numerical data are sufficiently
detailed to provide matching of pieces from the smoothness
condition, then such procedure allows to account for the main
scaling corrections. By this reason we use the usual and not
logarithmic coordinates.
\begin{figure}
\centerline{\includegraphics[width=3.1 in]{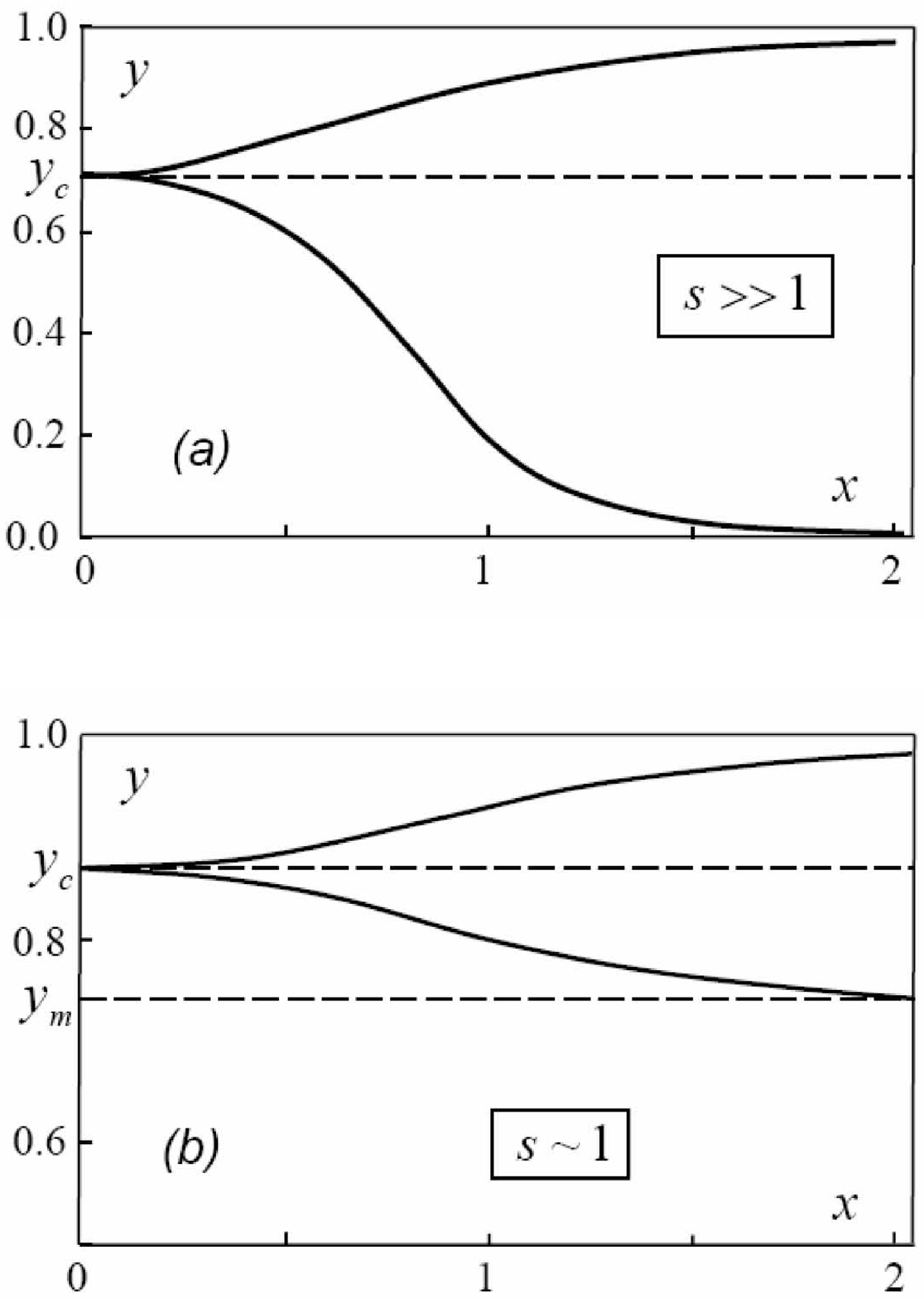}}
\caption{Examples of scaling dependencies  $y(x)$ for level
statistics in high dimensions: (a) for $s\gg 1$ ($u_0=0$) and  (b)
for $s\sim 1$ ($u_0=22.3$). The choice of parameters
$k_1=0.0652$, $B=0.230$ corresponds to numerical data for  $d=4$
\cite{15}. } \label{fig5}
\end{figure}

\begin{center}
{\bf 4. Level statistics.}
\end{center}

In the analysis of  level statistics  \cite{15},
the following combination is of the main interest,
$$
y=\sigma^2/\sigma_P^2 \,,
\eqno(10)
$$
where $\sigma$ is a root-mean-square fluctuation of the number of
levels $N$ in the energy interval $E=s \Delta$, where  $\Delta$ is
the mean level spacing in a finite system, and $\sigma_P$ is a
value of  $\sigma$ for the Poisson statistics. This quantity is
closely related with parameter $A$ in the asymptotics of the
distribution function $P(s)$
$$
P(s)\sim \exp\left(-As \right)\,,\qquad
A=\sigma_P^2/\sigma^2
\eqno(11)
$$
of the distance  $\omega=s\Delta$  between the nearest levels in the
large  $s$ limit. According to  \cite{15}, the quantity (10) is a
function of variable $x$,  which is defined as
$$
x=s^{-1/4}\,\frac{L}{\xi} \left(\frac{L}{a}
\right)^{(d-4)/4}, \qquad d>4 \,,
\eqno(12)
$$
$$
x=s^{-1/4}\,\frac{ L }{\xi} \frac{\left[\ln(\xi/a)\right]^{1/2} }
  {\left[\ln(L/a)\right]^{1/4}}     , \qquad d=4 \,.
\eqno(13)
$$
Dependence  $y(x)$ in the parametric form is given by equations
$$
y=\frac{\sigma^2}{\sigma_P^2}
 =k_1 u \ln  \frac{1+k_1  +k_1 u}{k_1 + k_1 u}\,,
$$
$$
\pm \,x^2 =
\frac{ (1+u)^{1/2} -B(u-u_0)}{(u-u_0)^{1/2}} \,,
\eqno(14)
$$
where the running variable  $u$ changes from $u_0$ till infinity.
Parameters  $B$ and $k_1$ are chosen according to the procedure
described in \cite{15}; parameter  $u_0$ account for a finiteness
of  $s$ and disappears for $s\to\infty$. The form of equations
(14) is the same for all  $d\ge 4$, while the choice of parameters
depends on  $d$.

Practically,  the following quantity is used as a  scaling
variable  \cite{12}
$$
J_0={\textstyle\frac{1}{2}}\langle s^2 \rangle =
{\textstyle\frac{1}{2}}\int_0^\infty s^2 P(s)\, ds \,,
\eqno(15)
$$
or another quantity  $\eta$ \cite{13}, closely related with it:
$$
\eta=\frac{J_0-J_{0W}}{J_{0P}-J_{0W}}=
     \frac{J_0-0.643}{0.357}  \,,
\eqno(16)
$$
where indices $W$ and $P$ mark the values of $J_0$ for the
Wigner--Dyson and Poisson statistics, $J_{0W}=0.643$, $J_{0P}=1$.
\begin{figure*}
\centerline{\includegraphics[width=6.6 in]{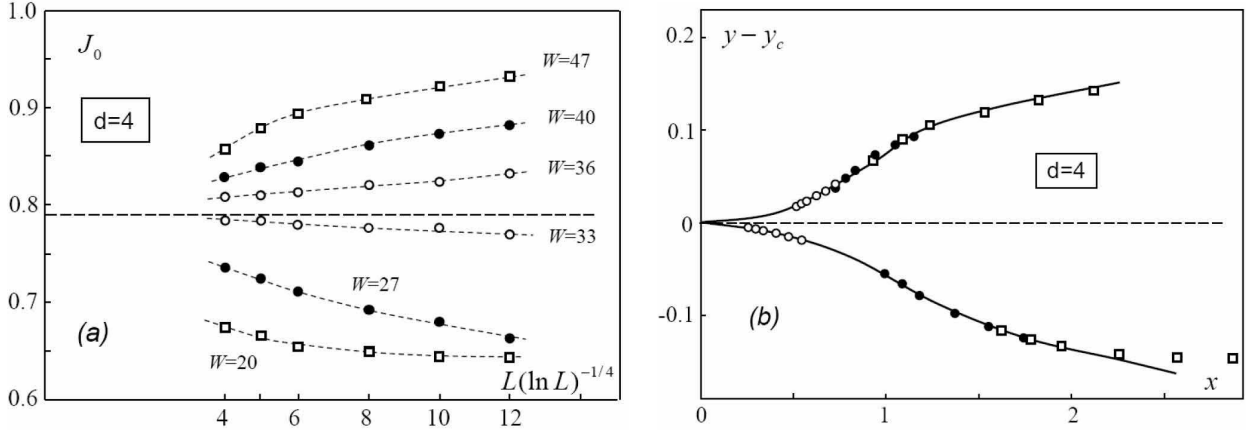}} \caption{(a)
Numerical data by Zharekeshev and Kramer for  $d=4$ (level
statistics), extracted from Fig.4 of paper  \cite{12} and
presented as a functional dependence on the "modified length"
$\mu(L)=L(\ln L)^{-1/4}$; (b) their comparison with the
theoretical scaling dependence (Fig.5,b). } \label{fig6}
\end{figure*}

Quantities $J_0$ and  $\eta$ are the regular functions of
variable  (10) for some $s\sim 1$ \cite{15}, so their
small deviations from the critical values are proportional
to each other:
$$
J_0-J_{0c} \sim \eta -\eta_c \sim y-y_c \,.
 \eqno(17)
 $$
Examples of dependencies  $y(x)$ for $s\to\infty$ (a) and $s\sim
1$ (b) are presented in Fig.5; they correspond to the critical
value  $A_c=1.4$ of parameter $A$ in (11), which is
specific for  $d=4$ \cite{12}.  Parameter  $u_0$ in
Fig.5,b is chosen so as to provide the approximate symmetry of
two branches, discovered in numerical experiments by
Zharekeshev and Kramer \cite{12} (Fig.6,a). These data can be
successfully matched with the theoretical dependence (Fig.6,b).

The critical value $y_c$ (Fig.5,a) tends to unity with
the increase of  $d$. Indeed, starting from the critical values
  $A_c=1.4$ ($d=4$) \cite{12}, $A_c=1.17$ ($d=5$) \cite{13},
  $A_c=1.13$ ($d=6$) \cite{13} and following the procedure of
  paper  \cite{15}, one can obtain   $y_c=0.714$ ($d=4$),
  $y_c=0.858$ ($d=5$), $y_c=0.885$ ($d=6$). This tendency
agrees with the theorem \cite{33},  that  level statistics
for the Bete lattice  (corresponding to $d=\infty$)
has the Poisson form even in the metallic phase.

With this observation, it is possible to  obtain the universal
scaling function for high dimensions. Suggesting $1-y\ll 1$, one
can expand the first equation (14) in  $1/k_1u$ and linearize the
right hand side of the second equation (14) near the critical
value  $u_c$; then
$$
y-y_c ={\rm const}\, F(x) \,,
\qquad F(x)=\frac{\pm x^2}{1 \pm x^2}\,,
\eqno(18)
$$
where an appropriate choice of the common scale along the $x$
axis is made; signs  $+$  and  $-$ corresponds to the upper
and lower  branches of function  $F(x)$ shown in Fig.7;
singularity at  $x=1$  is fictitious and lies beyond
the limits of applicability of
(18). According to  (17), deviations of $J_0$ and  $\eta$ from the
critical values are described by the same function.
\begin{figure}
\centerline{\includegraphics[width=3.1 in]{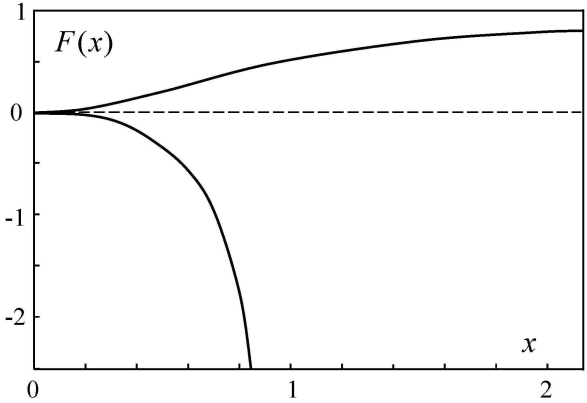}}
\caption{The universal scaling dependence
for level statistics, corresponding to high
dimensions. }
\label{fig7}
\end{figure}

Fig.8,a and Fig.9,a  illustrates the numerical data
by Garcia--Garcia and Cuevas  \cite{13} for $d=5$ and $d=6$
presented as a functional dependence on the "modified length"
$\mu(L)=L^{d/4}$;
they are in a good agreement with the universal scaling
function $F(x)$ (see Fig.8,b and Fig.9,b).
\begin{figure*}
\centerline{\includegraphics[width=6.6 in]{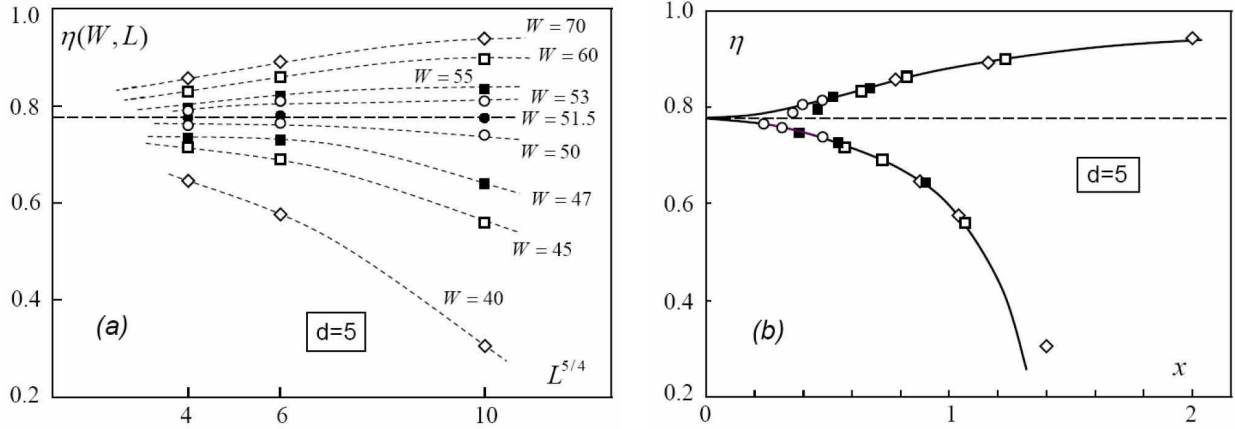}}
\caption{Numerical data by Garcia--Garcia and
Cuevas for $d=5$ (level statistics), extracted from
Fig.1 of paper  \cite{13} (a) and their comparison with the
theoretical scaling dependence shown in Fig.7  (b). }
\label{fig8}
\end{figure*}

In cases $d=5$ and  $d=6$ (in opposite to  $d=4$) corrections
related with a finiteness of $s$ are not essential. Indeed,
according to  \cite{12} the quantity  $J_0$ at  $d=4$ runs the
interval $(0.64, 0.79)$ along the lower branch, and the interval
$(0.79, 1.00)$ along the upper branch.  According to \cite{13},
these intervals are  $(0.64, 0.92)$,   $(0.92, 1.00)$ for $d=5$,
and $(0.64, 0.95)$, $(0.95, 1.00)$  for $d=6$.  If a choice of
$u_0$ provides the same proportion for the intervals $(y_m, y_c)$
and $(y_c, 1)$ in Fig.5,b, then $1-y_c\approx y_c-y_m$ for $d=4$
and $1-y_c\ll y_c-y_m$ for $d=5,\,6$.  In the latter case, the
difference of  $y_m$ from zero is practically not manifested in
the region of applicability for (18).
\begin{figure*}
\centerline{\includegraphics[width=6.6 in]{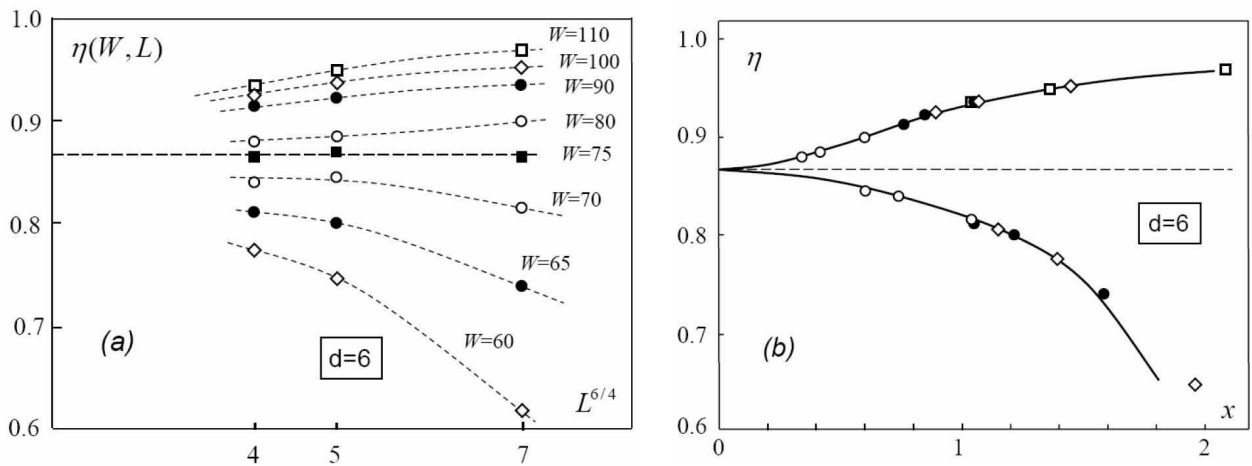}}
\caption{Numerical data by Garcia--Garcia and
Cuevas for $d=6$ (level statistics), extracted from
Fig.1 of paper  \cite{13} (a) and their comparison with the
universal scaling dependence (b). }
\label{fig9}
\end{figure*}

For $d>4$, equations (14) define scaling of the form
$$
y=\frac{\sigma^2}{\sigma_P^2}=
F\left( \frac{L^{d/4}}{\xi a^{(d-4)/4}} \right) \,,
\eqno(19)
$$
which for large  $\xi$ gives
$$
y=\tilde F\left( \tau \frac{L^{d/4\,\nu}}
{a^{(d-4)/4\,\nu}} \right) = y_c + C \tau L^{d/4\,\nu}
+\ldots \,,
\eqno(20)
$$
i.e. derivative  $y'_\tau$ at $\tau=0$ has a behavior
$L^{d/4\,\nu}$ instead  $L^{1/\nu}$ corresponding to scaling
of type (1). Hence, values of the exponents $\nu=0.84$ ($d=5$),
$\nu=0.78$ ($d=6$) obtained in \cite{13} in the framework of
(1) transform to  $\nu=0.67$ ($d=5$),
$\nu=0.52$ ($d=6$) if  (19) is used. Therefore, the results
of paper \cite{13} for  $d=5,\,6$ becomes close to $\nu=1/2$
simply in the result of transfer to a correct scaling relation.
Situation for  $d=4$ is analogous to that of Sec.\,3, i.\,e.
the main body of data corresponds to quasi-linear portions
$y\sim x \sim L (\ln L)^{-1/4}$ of the scaling curves, which
was interpreted in  \cite{12,13} as $L^{1/\nu}$ with
$\nu\approx 1$. Such situation is aggravated by the
accepted scheme
of treatment when a derivative over  $\tau=W-W_c$ is
determined by expansion over $W-W_c$ and fitting by a polynomial
of finite degree. In such procedure, the result is dominated by
experimental points remote of  $W_c$ and linearity in $x$
preserves even in  case, when the data close to  $W_c$
demonstrate an essential nonlinearity.

\begin{center}
{\bf 5. Conclusion}
\end{center}

The present paper suggests a new interpretation of existing
numerical data for the Anderson transition in high dimensions:
results for  $d=4,\,5$ obtained from scaling in
quasi-one-dimensional systems, and results for $d=4,\,5,\,6$
obtained from level statistics. Such reinterpretation is necessary
due to the absence of one-parameter scaling \cite{9} in high
dimensions, which is a consequence of nonrenormalizability of
theory. All indicated numerical data appear to be compatible with
theoretical scaling dependencies obtained from self-consistent
theory of localization by Vollhardt and W${\rm {\ddot o}}$lfle. It
supports the same tendency as was observed in preceding papers
\cite{14,15,19,20}: on the level of raw data, the Vollhardt and
W${\rm {\ddot o}}$lfle theory looks satisfactory,
while the opposite statements of the original papers are related
with ambiguity of the treatment procedure. It gives new arguments
in favor of the viewpoint
 \cite{17,18} that the Vollhardt and W${\rm {\ddot o}}$lfle
 theory predicts the exact critical behavior.


\end{document}